\documentclass[
aps,
prd,
twocolumn,
showpacs,preprintnumbers,nofootinbib,
amsmath,amssymb,
aps,
superscriptaddress,english]{revtex4-1}

\usepackage{graphicx}
\usepackage{dcolumn}
\usepackage{natbib,aas_macros}
\usepackage{color}
\usepackage{amssymb}
\usepackage{amsthm}
\usepackage{mathrsfs}
\usepackage{amsmath}
\usepackage{hyperref}
\usepackage{lineno}
\usepackage{units}
\usepackage{subcaption}
\usepackage{float}
\usepackage{multirow}
\usepackage{tabularx}
\usepackage[title]{appendix}

\usepackage{caption}
\definecolor{pink}{RGB}{255, 20, 147}

\definecolor{comment}{RGB}{166, 38, 164}

\usepackage{xparse}
\usepackage{xcolor}

\NewDocumentCommand{\codeword}{v}{%
\texttt{\textcolor{black}{#1}}%
}

\begin{document} 
\title{Probing resonant excitations in exotic compact objects via gravitational waves}

\author{Yasmeen Asali}
\email{ysa2106@columbia.edu}
\address{Department of Astronomy, Columbia University, 538 West 120th Street, New York, New York 10027, USA}
\address{Nikhef -- National Institute for Subatomic Physics, Science Park 105, 1098 XG Amsterdam, The Netherlands}

\author{Peter T. H. Pang}
\email{thopang@nikhef.nl}
\address{Nikhef -- National Institute for Subatomic Physics, Science Park 105, 1098 XG Amsterdam, The Netherlands}
\address{Department of Physics, Utrecht University, Princetonplein 1, 3584 CC Utrecht, The Netherlands}

\author{Anuradha Samajdar}
\address{Nikhef -- National Institute for Subatomic Physics, Science Park 105, 1098 XG Amsterdam, The Netherlands}
\address{Department of Physics, Utrecht University, Princetonplein 1, 3584 CC Utrecht, The Netherlands}

\author{Chris Van Den Broeck}
\address{Nikhef -- National Institute for Subatomic Physics, Science Park 105, 1098 XG Amsterdam, The Netherlands}
\address{Department of Physics, Utrecht University, Princetonplein 1, 3584 CC Utrecht, The Netherlands}

\begin{abstract}

Gravitational waves (GWs) from presumed binary black hole mergers are now being
detected on a regular basis with the Advanced LIGO and Advanced Virgo interferometers. 
Exotic compact objects (ECOs) have been proposed that differ from Kerr black holes, 
and which could leave an imprint upon the GW signal in a variety of 
ways. Here we consider excitations of ECOs during inspiral, which 
may occur when the monotonically increasing GW frequency matches a resonant 
frequency of an exotic object. This causes orbital energy to be taken away, 
leading to a speed-up of the orbital phase evolution. We show that resonances 
with induced phase shifts $\lesssim 10$ radians can be detectable with 
second-generation interferometers, using Bayesian model selection. 
We apply our methodology to detections in the GWTC-1 catalog from the first and second
observing runs of Advanced LIGO and Advanced Virgo, finding consistency with the
binary black hole nature of the sources. 

\end{abstract} 
\maketitle

\section{Introduction}
\label{sec:Introduction}

In the past several years, the Advanced LIGO observatories \cite{TheLIGOScientific:2014jea} together
with Advanced Virgo \cite{TheVirgo:2014hva} have been detecting gravitational wave (GW) 
signals from coalescing compact binaries on a regular basis. This includes a confirmed 
binary neutron star inspiral 
\cite{TheLIGOScientific:2017qsa,Soares-Santos:2017lru,Monitor:2017mdv,Cowperthwaite:2017dyu}, 
and more recently another possible binary neutron star \cite{Abbott:2020uma}, although 
most sources appear to have been binary black 
holes \cite{Abbott:2017xlt,Abbott:2016nmj,TheLIGOScientific:2016pea,Abbott:2017vtc,Abbott:2017gyy,LIGOScientific:2018mvr}. 
The detections made during the first and second observing runs are summarized in 
\cite{LIGOScientific:2018mvr}; the latter will be referred to as GWTC-1, for 
Gravitational Wave Transient Catalog 1. (For other detection efforts, see 
\cite{Nitz:2018imz,Venumadhav:2019tad,Venumadhav:2019lyq}.)

A number of alternatives to the Kerr black holes of classical general relativity have 
been proposed, called exotic compact objects (ECOs). For instance, if dark matter is composed
of fermionic particles then they may form star-like objects supported by degeneracy 
pressure: dark matter stars \cite{Kouvaris:2015rea}. Boson stars \cite{Liebling:2012fv} are 
macroscopic objects made out of scalar fields, as motivated by the discovery of the Higgs, 
cosmological inflation, axions as a solution of the strong CP problem, moduli in string theory, 
as well as dark matter. It has also been speculated that there may be gravastars \cite{Mazur:2003xz}:
objects with a so-called de Sitter core where spacetime is self-repulsive -- much like dark energy -- 
and held together by a shell of matter. As far as quantum gravity is concerned, fundamental 
considerations such as Hawking's information paradox have led some to postulate quantum 
modifications of the black holes of general relativity, such as firewalls \cite{Almheiri:2012rt} and 
fuzzballs \cite{Lunin:2001jy}. For an overview of the various ECOs that have been proposed in 
the literature, see \emph{e.g.}~\cite{Barack:2018yly}. 

When ECOs are part of a binary system that undergoes
coalescence, they can make their presence known through a variety of effects that may 
get imprinted upon the gravitational wave signal, and which would not be there in the 
case of ``standard" binary black holes. These include tidal effects 
\cite{Cardoso:2017cfl,Johnson-McDaniel:2018uvs}, dynamical friction as well as 
resonant excitations due to dark matter clouds in the vicinity of the objects 
\cite{Baumann:2018vus}, violations of the no-hair conjecture \cite{Carullo:2018sfu,Brito:2018rfr}, 
and gravitational wave ``echoes" following a merger  
\cite{Cardoso:2016rao,Cardoso:2016oxy,Cardoso:2017cqb,Cardoso:2019apo,Chen:2019hfg,Addazi:2019bjz,Cardoso:2017njb,Cardoso:2019rvt}.

Here we will focus on another possible signature of ECOs, namely resonant excitations 
during inspiral. Such effects have been well-studied in the context of neutron stars 
\cite{Kokkotas:1995xe,Lai:1993di,Shibata:1993qc,2007PhRvD..75d4001F,Yu:2016ltf}: as the
gravitational wave frequency increases monotonically, at one or more points in time it can 
become equal to an internal resonant frequency of a compact object. The resulting
excitation takes away part of the orbital energy, leading to a speed-up 
of the orbital motion, which in turn affects the phasing of the gravitational 
wave signal. As pointed out in \cite{Macedo:2013qea,Macedo:2013jja}, for inspiraling boson stars 
the gravitational wave signature of such effects can potentially be detected. 

Thus, 
it is natural to ask whether there is any sign of resonant excitations in the signals 
of the presumed binary black hole coalescences of GWTC-1. In this paper we develop a 
concrete data analysis framework to search for these signatures in data from Advanced LIGO
and Advanced Virgo, and apply it on events from GWTC-1. In particular, in 
Sec.~\ref{sec:methodology} we outline
our basic set-up for the effect of resonant excitations on the gravitational wave phase, together
with the methodology to search for resonances in gravitational wave signals. 
This is then applied to simulated signals in Sec.~\ref{sec:simulations}, where we assess
the detectability of the effect. The signals in GWTC-1 are analyzed in Sec.~\ref{sec:gwtc-1}. 
Finally, Sec.~\ref{sec:conclusions} provides a summary and future directions.

\section{Methodology}
\label{sec:methodology}

\subsection{Imprint of resonant excitations on the gravitational wave phase}

Our model of how resonant excitations modify the gravitational wave signal from inspiraling 
ECOs will be based on that of Flanagan et al.~\cite{2007PhRvD..75d4001F}; the context of that 
work was resonant r-modes in binary neutron star inspiral, but the basic assumptions carry
over to the case at hand. For simplicity, let us begin by assuming that only one of the 
two inspiraling objects undergoes a resonance, at some time $t_0$. The excitation takes away
part of the orbital energy, causing the gravitational wave phase $\Phi(t)$ to undergo an apparent 
advance in time $\Delta t$ relative to the point particle inspiral phase $\Phi_{\rm pp}(t)$:
\begin{equation}
\Phi(t) =
\begin{cases} 
\Phi_{\rm pp}(t) & \text{if } t  < t_{0},  \\ 
\Phi_{\rm pp}(t +\Delta t) - \Delta\Phi & \text{if } t \ge t_{0},
\end{cases}
\label{eq:phase-cases}
\end{equation}
where the phase shift $\Delta \Phi$ is such that $\Phi(t)$ remains continuous at $t = t_0$. 
Assuming $\Delta t$ to be sufficiently small, we may write
\begin{equation}
\Delta \Phi = \dot{\Phi}_{\rm pp}(t_{0})\Delta t. 
\label{DPhi}
\end{equation}
Expanding $\Phi(t)$ in Eq.~(\ref{eq:phase-cases}) to linear order in $\Delta t$, we then obtain
\begin{equation} 
\Phi(t) = \Phi_{\rm pp}(t) + \dot{\Phi}_{\rm pp}(t)\Delta t - \dot{\Phi}_{\rm pp}(t_{0})\Delta t. 
\end{equation}
The instantaneous gravitational wave frequency is $\omega = \dot{\Phi}$; using this and 
Eq.~(\ref{DPhi}) leads to 
\begin{equation}
\Phi(t) = \Phi_{\rm pp}(t) + \theta(t - t_0)\, \bigg[\frac{\omega(t)}{\omega(t_{0})} - 1\bigg]\Delta\Phi,
\label{eq:phase-shift-time-domain}
\end{equation}
with $\theta(t - t_0)$ the usual step function. 
(Clearly we are assuming that the resonant excitation is of sufficiently short duration
so as to be near-instantaneous. At least in the boson star examples of 
\cite{Macedo:2013jja,Macedo:2013jja} this is a reasonable approximation, but it may 
not be typical.) 
In the stationary phase approximation \cite{Sathyaprakash:1991mt}, this implies that 
the phase $\phi(f)$ of the frequency domain waveform becomes
\begin{equation}
\phi(f) = \phi_{\rm pp}(f) + \theta(f - f_0)\,\left[\frac{f}{f_{0}} - 1\right]\Delta\Phi,
\label{eq:phase-shift-freq-domain}
\end{equation}
where $\phi_{\rm pp}(f)$ is the point particle phase in the Fourier domain, and $f_0$  
the frequency at which the resonance occurs. 

In practice, both objects in the binary system may experience resonant excitation; moreover, 
an individual object may be subject to several resonant excitations at different frequencies during
the time the signal is in the detectors' sensitive frequency band \cite{Macedo:2013jja,Macedo:2013jja}. 
In order to keep the data 
analysis problem tractable in terms of computational requirements as well as the 
dimensionality of parameter space, in this work we will allow for up to two main instances of resonance, 
with associated frequencies $f_{01}$ and $f_{02}$, assuming other resonances to have
a negligible effect. The frequency domain phase then becomes
\begin{eqnarray}
\phi(f) &=& \phi_{\rm pp}(f) \nonumber\\
&& + \, \theta(f - f_{01}) \left[\frac{f}{f_{01}} - 1\right]\Delta\phi_{01} \nonumber\\
&& + \, \theta(f - f_{02}) \left[\frac{f}{f_{02}} - 1\right]\Delta\phi_{02}.
\label{eq:full-phase-shift}
\end{eqnarray}
In what follows, we will assume that the values of resonance frequencies are ordered such that
$f_{01} < f_{02}$.

In order for resonances to be observable, it is necessary that (a) the cumulative dephasing
with respect to the point particle case is sufficiently large (for second-generation 
detectors this can be taken to mean larger than $\sim 1$ radian), and (b) the resonance frequencies 
are within the detectors' sensitive frequency band. As shown in \cite{Cardoso:2019nis}, it is hard
to meet both of these criteria simultaneously for ECOs whose horizon modification scale 
is microscopic (as would be the case for e.g.~fuzzballs \cite{Lunin:2001jy}). On the other hand, 
the analysis of \cite{Macedo:2013qea,Macedo:2013jja} indicates that for boson stars, it is possible to 
satisfy both conditions at the same time.

\subsection{Bayesian analysis}
\label{sec:parameter-estimation}

Our expression (\ref{eq:full-phase-shift}) for the phase in the presence of resonances leads to 
a Fourier domain 
waveform model $\tilde{h}_{\rm ECO}(f)$ (which is discussed in more detail below), and this in turn defines a Bayesian 
hypothesis $H_{\rm ECO}$ which states that
resonances took place in a given coalescence event. This can be compared with the hypothesis $H_{\rm BBH}$ stating
that no resonances took place; the associated waveform model $\tilde{h}_{\rm BBH}(f)$ just describes the signal from 
binary black hole coalescence. Given a hypothesis $H$, data $d$, and whatever background information $I$ we possess, 
a Bayesian evidence is obtained through
\begin{equation}
p(d | H, I) = \int d\bar\theta\,p(d | H, \bar\theta, I)\,p(\bar\theta | H, I).
\end{equation}
The integral is over the parameters $\bar{\theta}$ (masses, spins, possible resonance frequencies and phase shifts, ...)
that the waveform $\tilde{h}(\bar{\theta}; f)$ depends on. $p(\bar\theta | H, I)$ is the prior density, and the likelihood 
$p(d | H, \bar\theta, I)$ is given by \cite{Veitch:2009hd}
\begin{equation}
p(d | H, \bar\theta, I) \propto \exp\left[-\langle d - h(\bar\theta) | d - h(\bar\theta) \rangle/2 \right],
\label{likelihood}
\end{equation}
where the noise-weighted inner product $\langle\,\cdot\,|\,\cdot\,\rangle$
is defined in terms of the noise power spectral density $S_n(f)$:
\begin{equation}
\langle a | b \rangle \equiv 4\Re \int_{f_{\rm low}}^{f_{\rm high}} df\,\frac{\tilde{a}^\ast(f)\,\tilde{b}(f)}{S_n(f)},
\end{equation}
with $f_{\rm low}$ and $f_{\rm high}$ respectively the lower cut-off frequency of a detector and the ending frequency 
of a given signal. This enables us to compute the ratio of evidences, or Bayes factor, for the hypotheses $H_{\rm ECO}$
and $H_{\rm BBH}$:
\begin{equation}
\mathcal{B}^{\rm ECO}_{\rm BBH} \equiv \frac{p(d | H_{\rm ECO}, I)}{p(d | H_{\rm BBH}, I)}.
\end{equation}

If for a given gravitational wave signal the (log) Bayes factor $\log \mathcal{B}^{\rm ECO}_{\rm BBH}$ is high, then this 
may be indicative of resonances having occurred. However, also noise artefacts can cause $\log \mathcal{B}^{\rm ECO}_{\rm BBH}$
to be elevated. In order to establish a statistical significance, we add a large number of simulated 
binary black hole signals to the detector noise and compute the log Bayes factor for all of them, leading to
a so-called background distribution $\mathcal{P}_{\rm BBH}(\log \mathcal{B}^{\rm ECO}_{\rm BBH})$. Given a real signal with a particular value for 
$\log \mathcal{B}^{\rm ECO}_{\rm BBH}$, the associated \emph{false alarm probability} (FAP) is given by
\begin{equation}
\mbox{FAP} = \int_{\log \mathcal{B}^{\rm ECO}_{\rm BBH}}^\infty \mathcal{P}_{\rm BBH}(x)\,dx.
\end{equation}

Next, consider a large number of simulated signals containing resonant effects, with given ranges for parameters
like masses, resonance frequencies, and induced phase shifts for the component objects. Let the distribution of 
$\log \mathcal{B}^{\rm ECO}_{\rm BBH}$ for these signals be $\mathcal{P}_{\rm ECO}(\log \mathcal{B}^{\rm ECO}_{\rm BBH})$. 
Given a threshold $p_{\rm th}$ for the 
false alarm probability, the \emph{efficiency} in uncovering the resonant effects is defined as 
\begin{equation}
\epsilon = \int_{\log \mathcal{B}_{\rm th}}^\infty \mathcal{P}_{\rm ECO}(y)\,dy, 
\label{efficiency}
\end{equation}
where the threshold $\log \mathcal{B}_{\rm th}$ on the log Bayes factor is obtained through
\begin{equation}
p_{\rm th} = \int_{\log \mathcal{B}_{\rm th}}^\infty \mathcal{P}_{\rm BBH}(x)\,dx.
\label{threshold}
\end{equation}

Apart from hypothesis testing we also measure the parameters associated with a hypothesis. Using the likelihood
function defined in Eq.~(\ref{likelihood}), a joint posterior density function for all the parameters
is obtained through Bayes' theorem:
\begin{equation}
p(\bar\theta | H, d, I) = \frac{p(d | H, \bar\theta, I)\,p(\bar\theta | H, I)}{p(d | H, I)}.
\label{PE}
\end{equation}
Finally, the one-dimensional posterior density function for a given parameter is obtained by 
integrating out all the other parameters.

The baseline of the waveform model was taken to be the frequency domain 
inspiral-merger-ringdown approximant IMRPhenomPv2  
\cite{Hannam:2013oca,Husa:2015iqa,Khan:2015jqa}, and modifications arising from resonances 
were added on top
of that; in particular, the phase was changed according to Eq.~(\ref{eq:full-phase-shift}). 
Priors for $\Delta\phi_{01}$ and $\Delta\phi_{02}$ were chosen to be uniform in $[0, 100]$. Those for  
$f_{01}$, $f_{02}$ were taken to be uniform in the interval 
$[20, 440]$ Hz, where the lower limit of the range is the detectors' $f_{\rm low}$ and 
the upper limit corresponds to the innermost
stable circular orbit (ISCO) for a 
total mass of $M = 10\, M_\odot$. For sources with a higher total mass 
this implies that our analyses will in practice 
also be searching for non-standard effects in the phase past the end of 
inspiral. In principle we could have restricted $f_{01}$, $f_{02}$ to be below
the ISCO frequency; however, allowing for an extended range has the benefit 
that we can be sensitive to more general departures from BBH behavior in the 
inspiral-merger-postmerger phase evolution than just resonant excitations of the component objects 
during inspiral. 

Finally, the software implementation of our methodology was based on the 
LIGO Algorithm Library Suite (\texttt{LALSuite}); the likelihood calculation was performed using the 
nested sampling algorithm in the \texttt{lalinference} package of 
LALSuite \cite{Veitch:2009hd,Veitch:2014wba}. 

\begin{figure*}
    \centering
    \includegraphics[width=\textwidth]{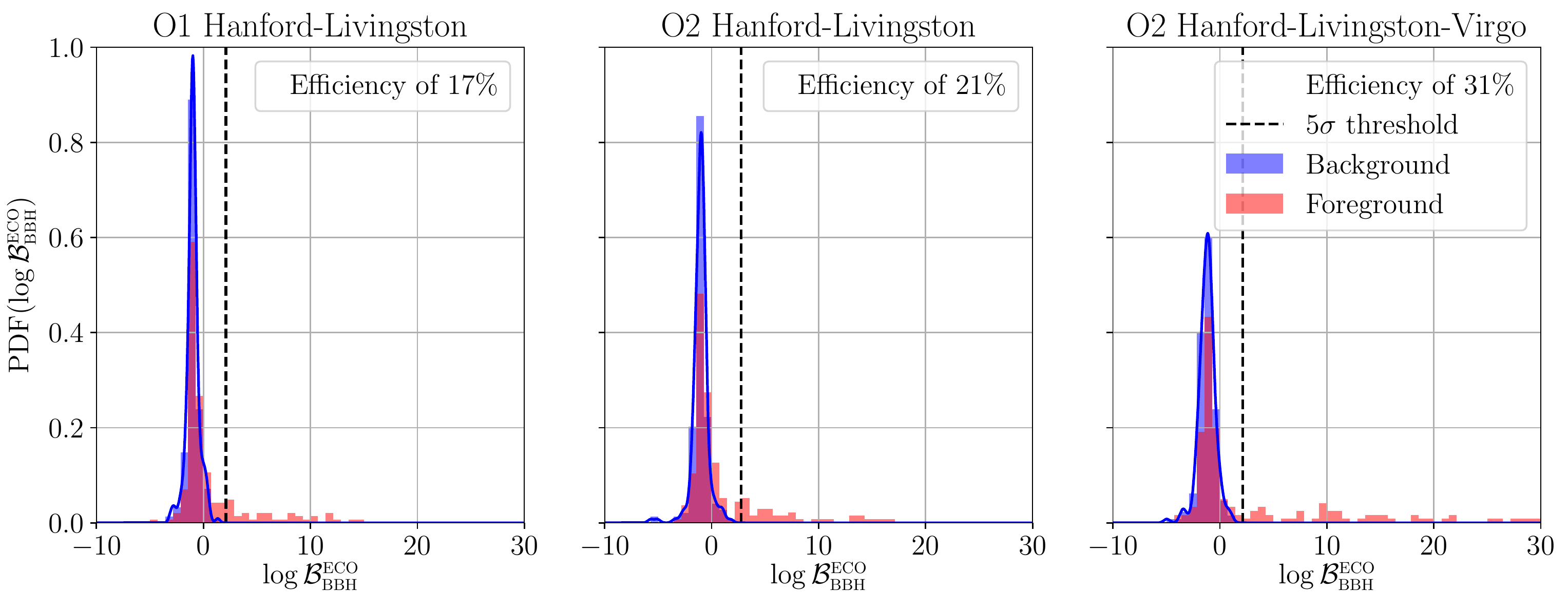}
    \caption{Distributions of log Bayes factors for the ECO hypothesis over the BBH hypothesis 
    for BBH injections (blue) and ECO injections (red), with parameter ranges as described in 
    the main text. For the BBH injections, we also show Gaussian KDE fits 
    (the smooth curves) to the background distribution, with respect to which 
    a $5\sigma$ threshold for detectability of resonances is established (the dashed vertical lines). 
    The left panel shows results for injections in O1 data from LIGO Hanford 
    and LIGO Livingston; the middle panel uses O2 data where only the two LIGO detectors 
    where active; and the right panel is for O2 when the two LIGO detectors as well as 
    Virgo were on. For the chosen $5\sigma$ threshold,  
    the log Bayes factor distributions for ECOs lead to efficiencies of, respectively, 
    17\%, 21\%, and 31\%.}
    \label{fig:background-foreground}
\end{figure*} 

\begin{figure}[t]
    \includegraphics[width=0.5\textwidth]{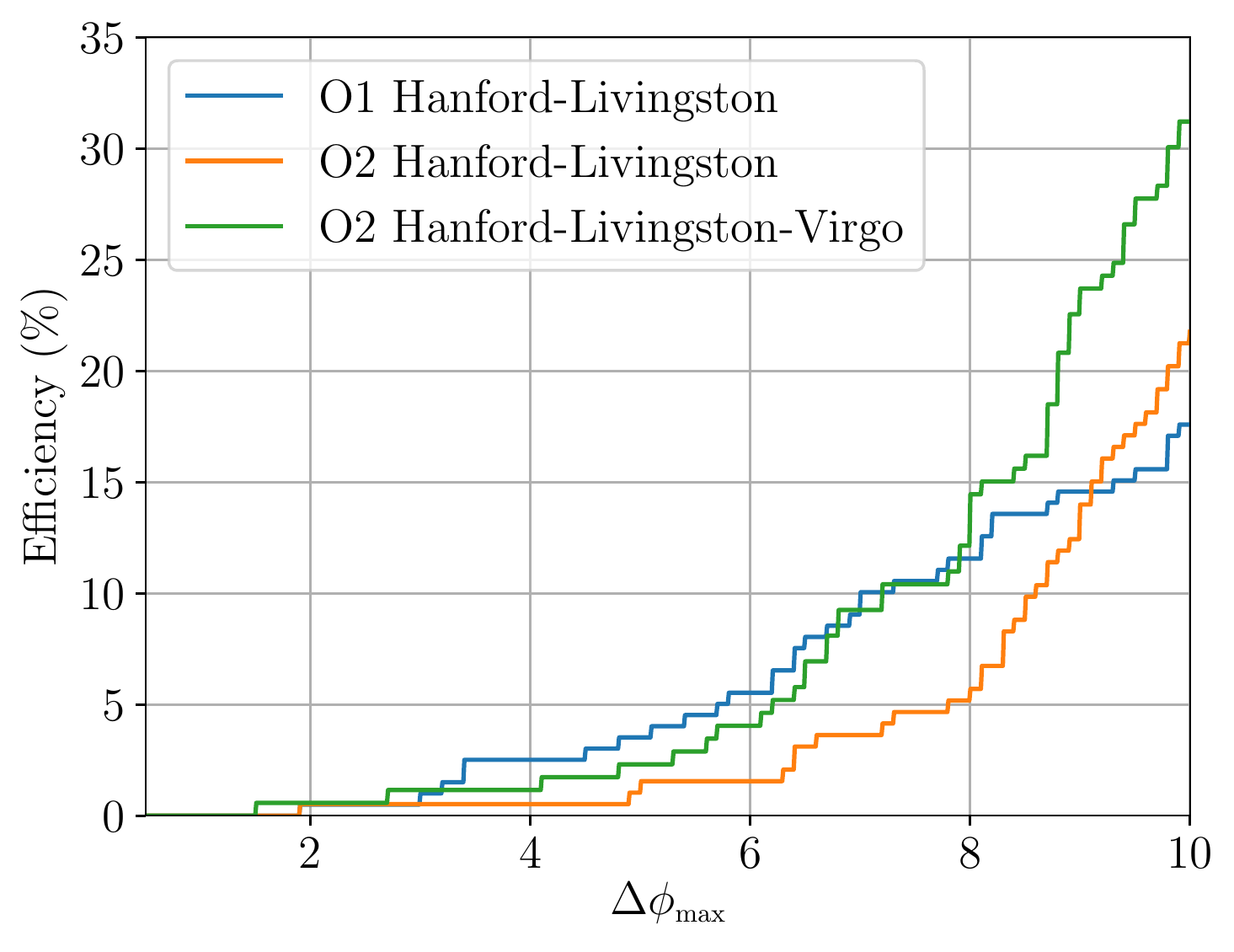} 
    \caption{Efficiencies with respect to a $5\sigma$ threshold as function of 
    the size of the phase shifts associated with resonances. We consider subsets of
    foreground samples in which the $\Delta\phi_{0i}$, $i = 1, 2$ do not exceed some given 
    $\Delta\phi_{\rm max}$, 
    and progressively increase this maximum value.} 
    \label{fig:efficiency-trend}
\end{figure}

\begin{figure}[t]
    \includegraphics[width=0.5\textwidth]{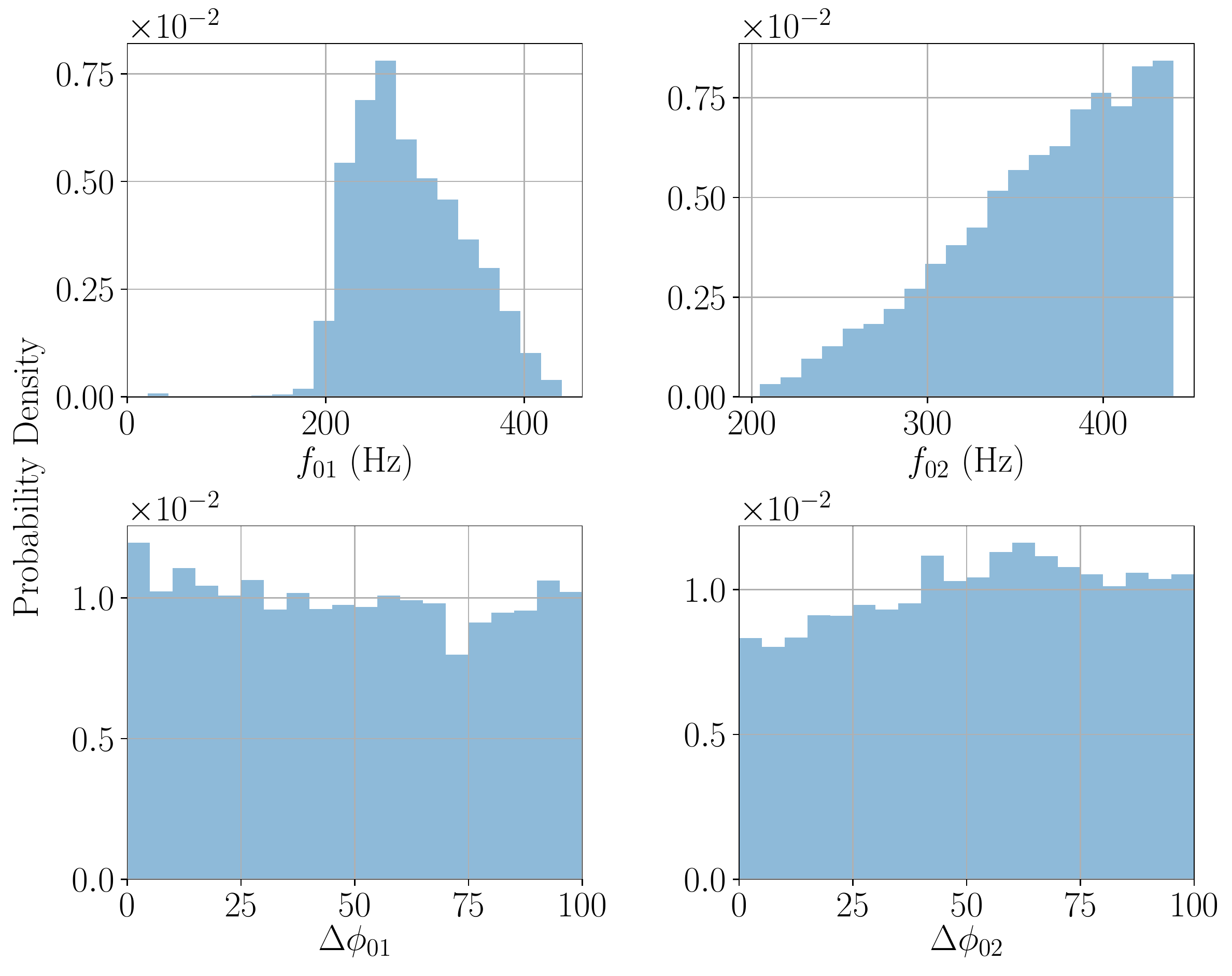}
    \caption{Posterior density functions for $f_{01}$, $f_{02}$, $\Delta\phi_{01}$, 
    and $\Delta\phi_{02}$, in a case where no resonances are present and the signal corresponds
    to a BBH. For this event the inspiral ended at $f_{\rm ISCO} = 94.7$ Hz; the posteriors
    for $f_{01}$ and $f_{02}$ mainly have support for frequencies well above that, 
    and the $\Delta\phi_{01}$, $\Delta\phi_{02}$ distributions largely return the prior.} 
    \label{fig:PE_BBH}
\end{figure}

\begin{figure}[t]
    \includegraphics[width=0.5\textwidth]{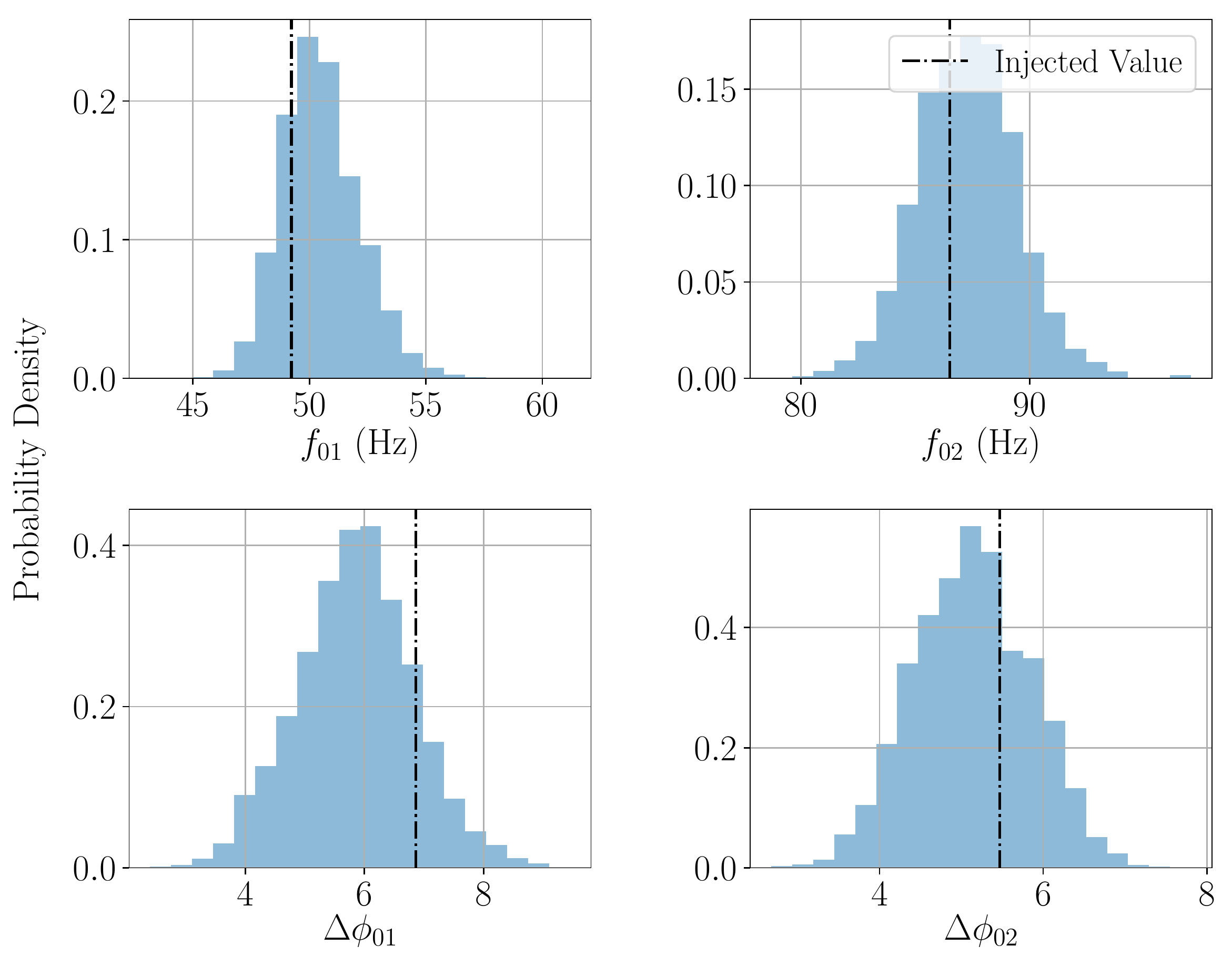}
    \caption{Posterior density functions for an example ECO with two resonances during 
    inspiral; the dashed-dotted vertical lines indicate the true values 
    of $f_{01}$, $f_{02}$, $\Delta\phi_{01}$, and $\Delta\phi_{02}$. In this case 
    $\log\mathcal{B}^{\rm ECO}_{\rm BBH} = 19.98$, \emph{i.e.}~above our $5\sigma$
    detection threshold for resonances.} 
    \label{fig:PE_2res}
\end{figure}

\begin{figure}[t]
    \includegraphics[width=0.5\textwidth]{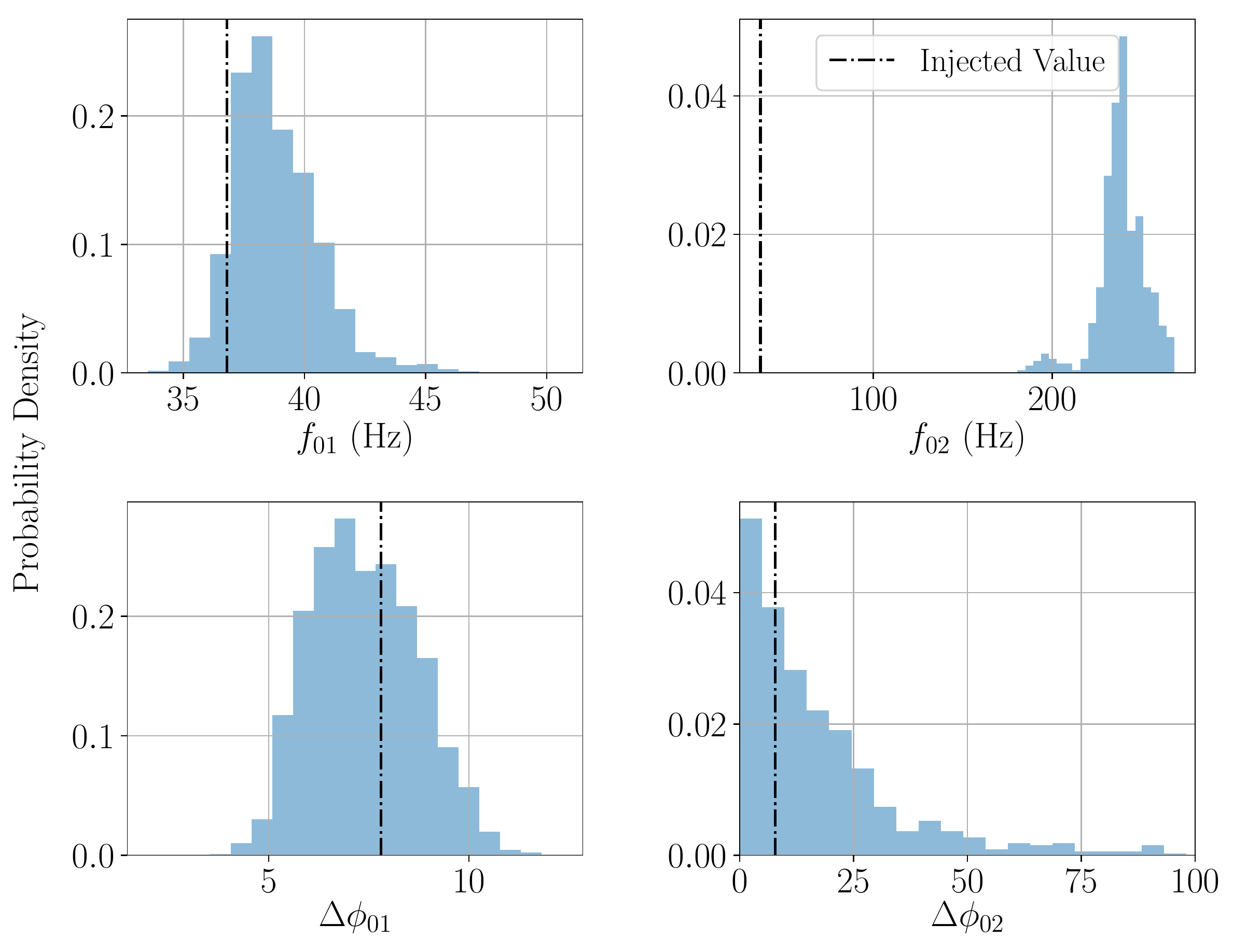}
    \caption{Posterior density functions for an example ECO with \emph{only one} resonance.  
    The dashed-dotted vertical lines indicate the true values of frequency and phase shift. 
    The posteriors for $f_{01}$ and $\Delta\phi_{01}$ capture these reasonably well, 
    but the posterior for $f_{02}$ again has support at values much above $f_{\rm ISCO} = 45.4$ Hz, similar
    to the BBH case of Fig.~\ref{fig:PE_BBH}. Also, the $\Delta \phi_{02}$ distribution is consistent with 0. In this example 
    $\log\mathcal{B}^{\rm ECO}_{\rm BBH} = 10.18$, well above the $5\sigma$ threshold.} 
    \label{fig:PE_1res}
\end{figure}

\begin{figure*}
    \centering
    \includegraphics[width=\textwidth]{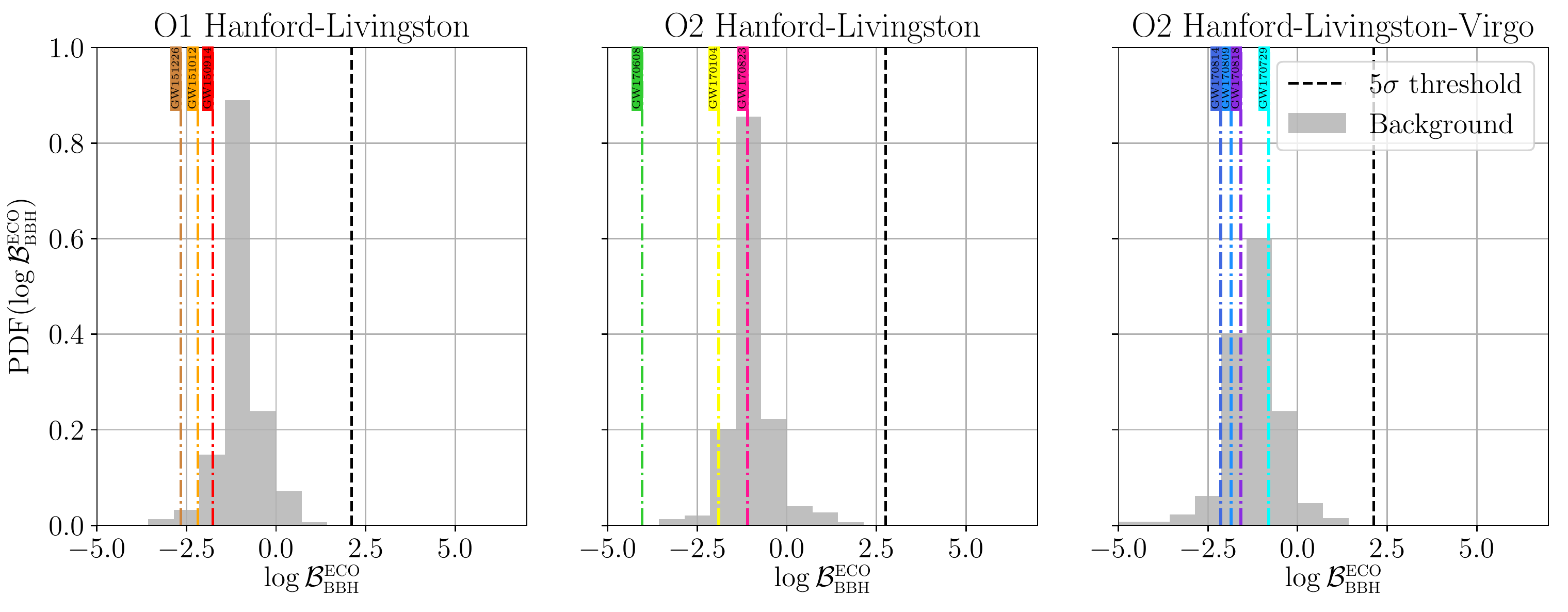}
    \caption{Log Bayes factors for the presumed binary black hole events of
    GWTC-1 (vertical dashed-dotted lines). For reference we also 
    show the BBH background distributions of Fig.~\ref{fig:background-foreground}
    again distinguishing between the case of two detectors in O1 (left), the
    two LIGO detectors in O2 (middle), and the two LIGOs together with Virgo in 
    O2 (right); vertical dashed lines again indicate $5\sigma$ significance thresholds
    as in Fig.~\ref{fig:background-foreground}. 
    It will be clear that for none of the GWTC-1 events, observable resonances are 
    present in any statistically significant way.}
    \label{fig:background-gwtc1}
\end{figure*}

\begin{table}[]
\begin{tabular}{| c| c| c | c |}
\hline
Sensitivity and IFOs            & Event   & $\log \mathcal{B}^{\textrm{\tiny{ECO}}}_{\textrm{\tiny{BBH}}}$  & FAP         \\ \hline
\multirow{3}{*}{O1 HL}   & GW150914 & -1.76 & 0.94   \\ 
                         & GW151012 & -2.18 & 0.97   \\ 
                         & GW151226 & -2.66 & 0.98   \\ \hline
\multirow{3}{*}{O2 HL}   & GW170104 & -1.90 & 0.96   \\ 
                         & GW170608 & -4.04 & 1.00   \\ 
                         & GW170823 & -1.09 & 0.59  \\ \hline
\multirow{4}{*}{O2 HLV}  & GW170729 & -0.81  & 0.25  \\ 
                         & GW170809 & -1.85  & 0.85   \\  
                         & GW170814 & -2.14  & 0.93  \\   
                         & GW170818 & -1.58 & 0.75     \\ \hline
\end{tabular}
\caption{Values of log Bayes factors for the GWTC-1 events, together with false alarm probabilities
with respect to the background distributions computed for the three kinds of data sets.}
\label{tab:logB}
\end{table}

\begin{figure}[t]
    \includegraphics[width=0.5\textwidth]{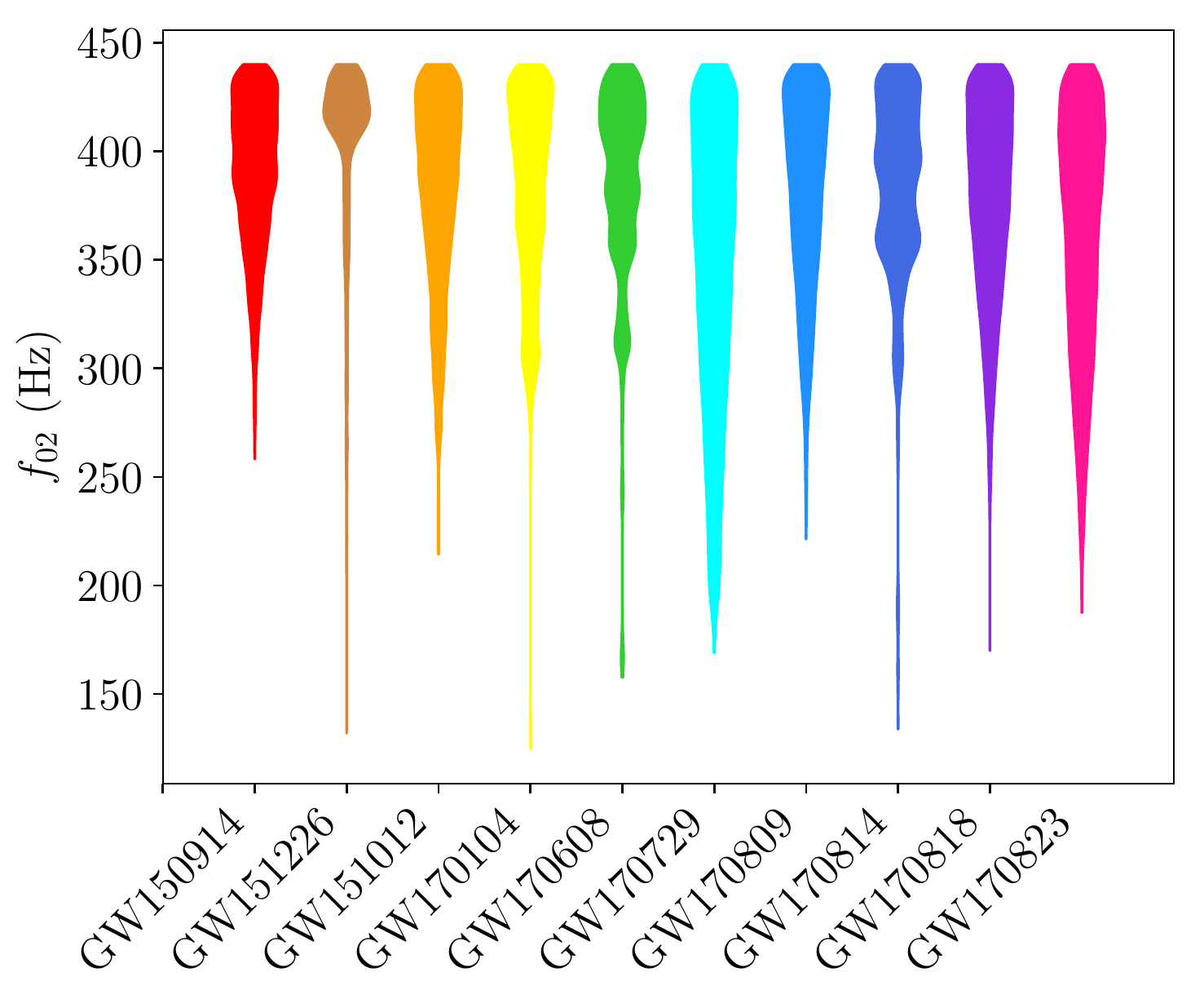} 
    \includegraphics[width=0.5\textwidth]{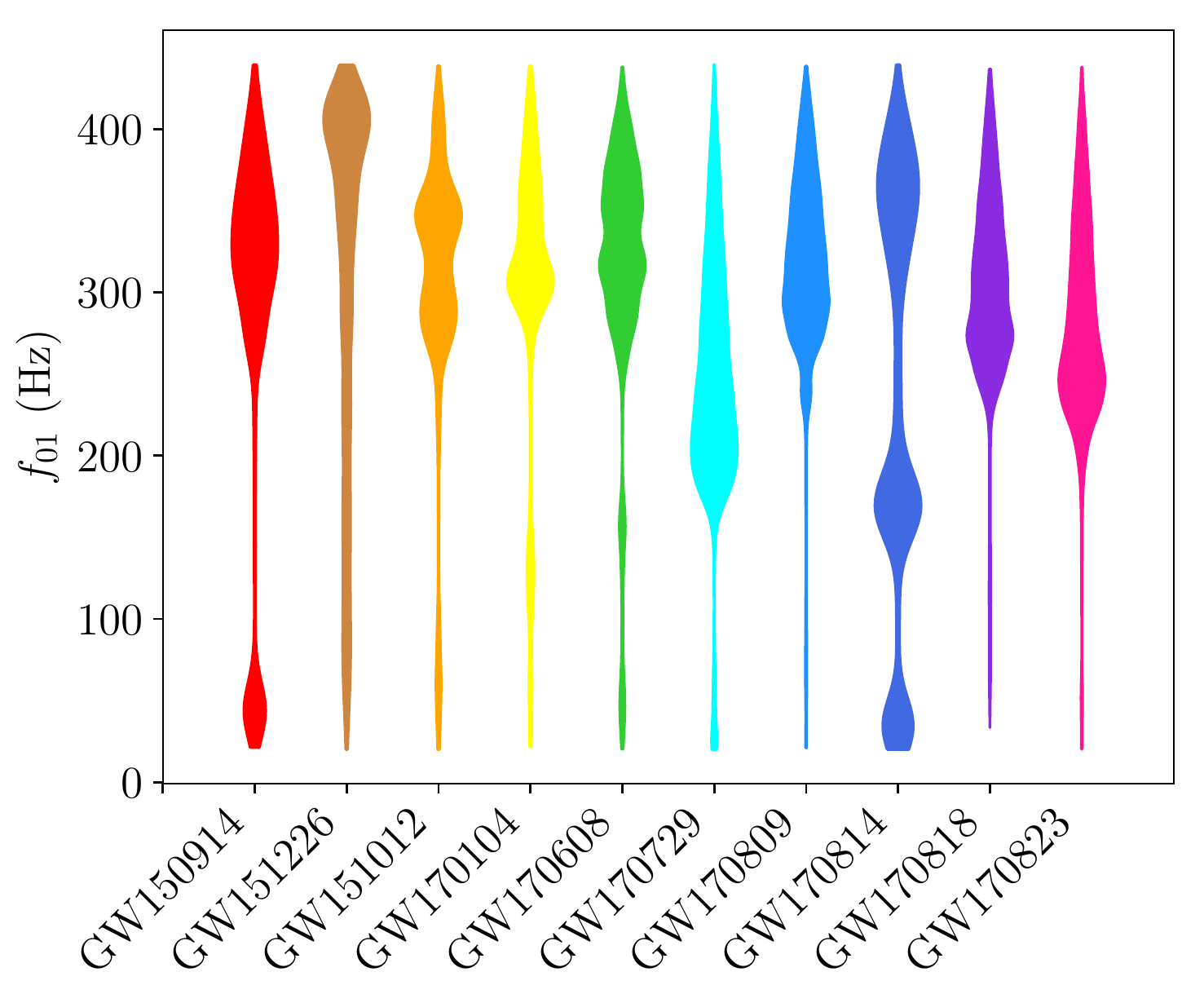}
    \caption{Posterior density functions for the  
    resonance frequencies $f_{02}$ (top) and $f_{01}$ (bottom), for each of the GWTC-1 events.} 
    \label{fig:freq}
\end{figure}

\begin{figure}[t]
    \includegraphics[width=0.5\textwidth]{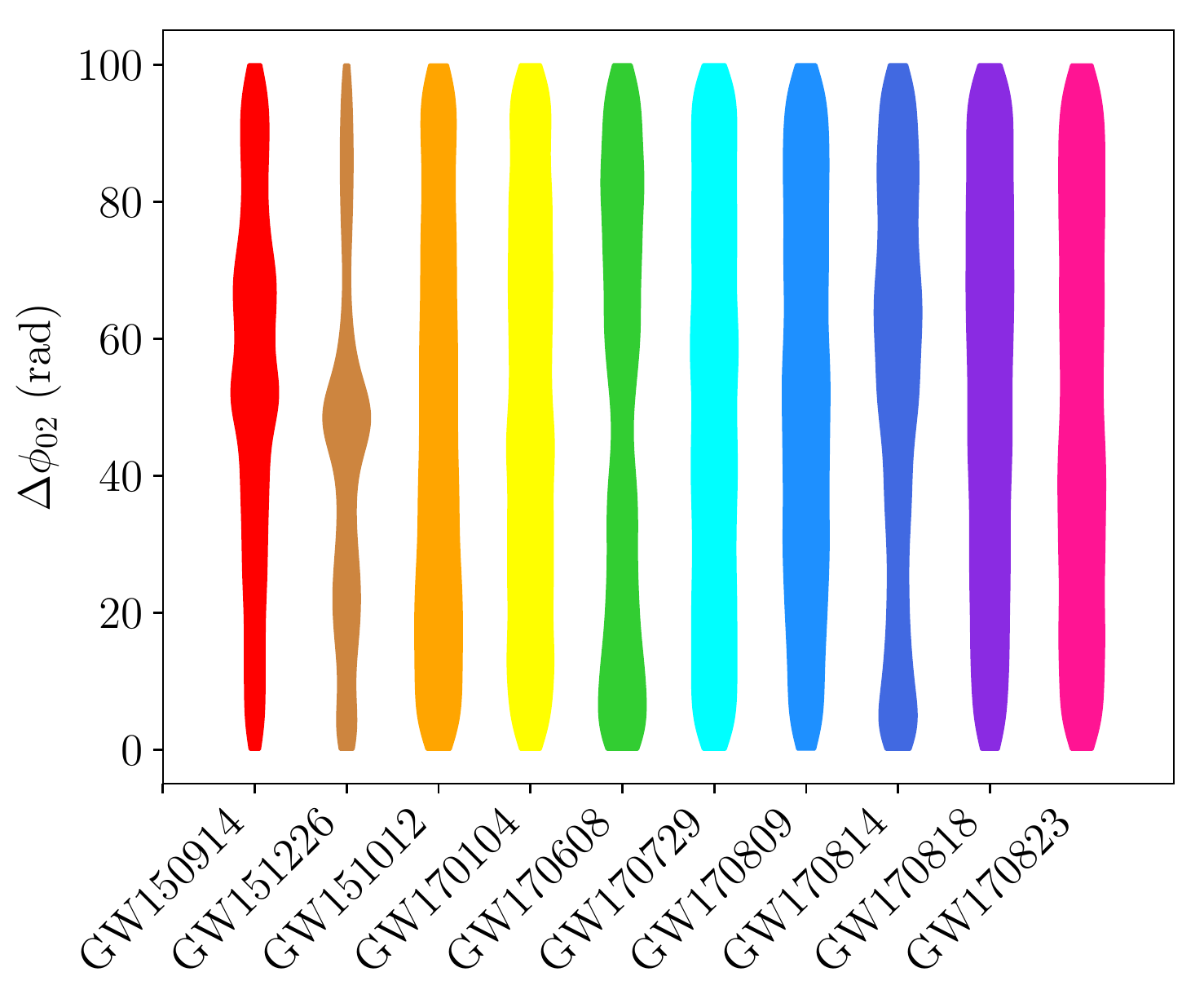}
    \includegraphics[width=0.5\textwidth]{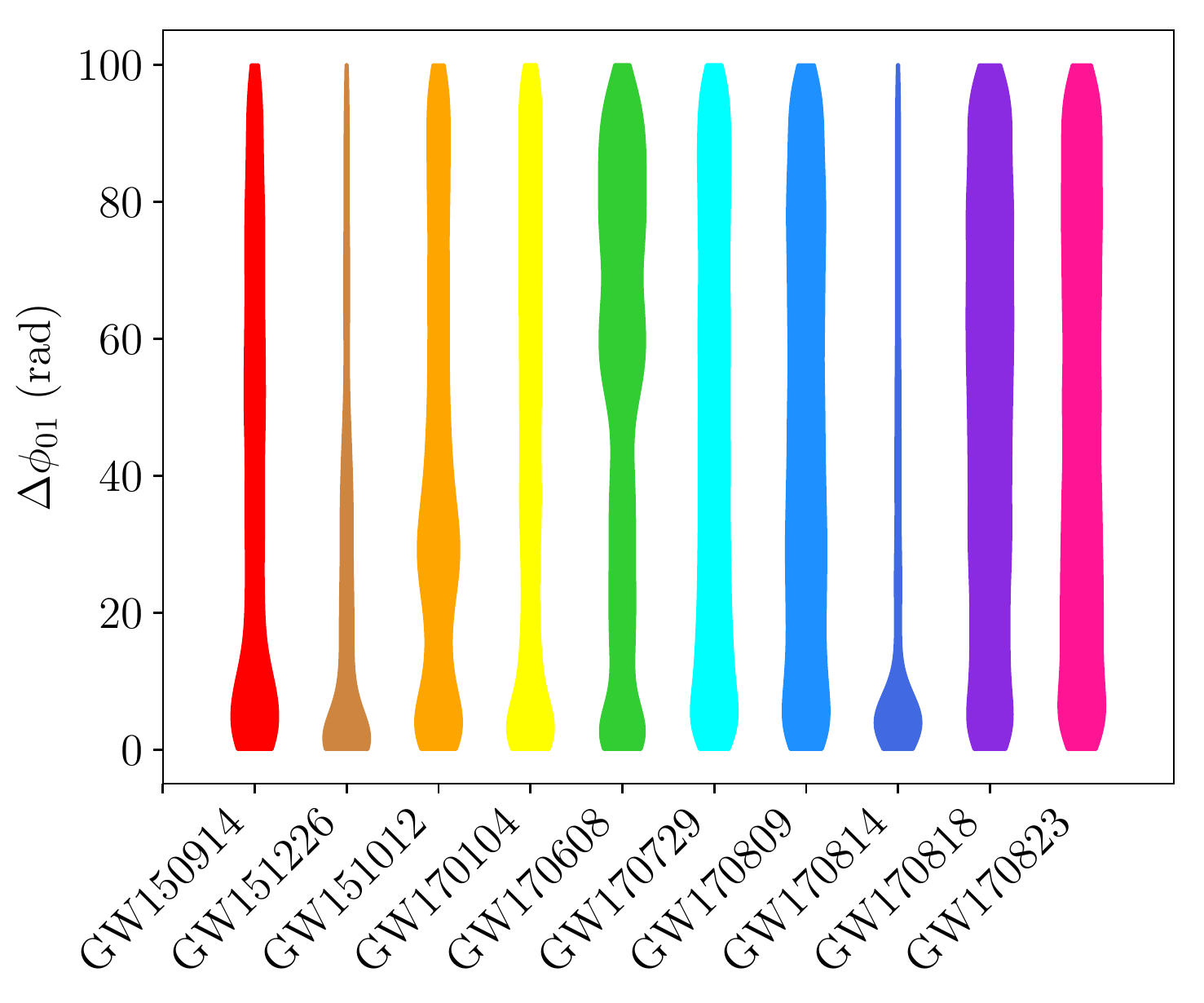}
    \caption{Posterior density functions for the phase shifts $\Delta\phi_{02}$ (top) and 
    $\Delta\phi_{01}$ (bottom), for each of the 
    GWTC-1 events.} 
    \label{fig:phase}
\end{figure}

\section{Simulations}
\label{sec:simulations}

\subsection{Measurability of resonance effects in the O1 and O2 observing runs}

First we want to gain some basic insight into the size of the resonance effects that could be measurable 
with Advanced LIGO and Advanced Virgo. To this effect we compute $\log \mathcal{B}^{\rm ECO}_{\rm BBH}$
for two sets of simulated signals, or \emph{injections}, in LIGO-Virgo data from the first
two observing runs (O1 and O2) \cite{LIGOScientific:2018mvr,O1_open_data,O2_open_data, Vallisneri_2015}, with one set corresponding to 
BBHs and the other to ECOs. 
For both sets, component masses are drawn uniformly from $m_1, m_2 \in [5, 70]\,M_\odot$, but 
total masses are restricted to $m_1 + m_2 \in [15, 110]\,M_\odot$, consistent
with the mass estimates for the BBH-like events in GWTC-1 \cite{LIGOScientific:2018mvr}. 
The latter leads to a maximum  
ISCO frequency of $f_{\rm ISCO, max} = 293$ Hz, and a median 
of $f_{\rm ISCO, median} = $ 83 Hz.  
Sources are distributed uniformly in volume, with a lower cut-off on the network signal-to-noise
ratio of $\mbox{SNR} \geq 8$. 

In the set of injections corresponding to ECOs, the induced phase shifts due 
to resonances are taken to be uniform in $\Delta\phi_{01}, \Delta\phi_{02} \in (0, 10]$ rad, and 
possible resonance frequencies are chosen uniformly in $f_{01} \in [20,50]$ Hz, $f_{02} \in [20, 100]$ Hz. 
We only implement phase modifications for frequencies below $f_{\rm ISCO}$. We 
reject injections for which both $f_{01}$ and $f_{02}$ are above $f_{\rm ISCO}$; of those remaining,     
59\% will exhibit only one resonance during inspiral, and the rest will have two of them. 
Finally, spin 
directions are chosen to be uniform on the sphere, with dimensionless spin magnitudes in 
the interval $[0, 1)$. 

Results are shown in Fig.~\ref{fig:background-foreground}, for 200 BBH injections and
an equal number of ECO injections. Distributions of 
$\log \mathcal{B}^{\rm ECO}_{\rm BBH}$ are displayed separately for the case of two LIGO detectors 
active in O1, two LIGO detectors active in O2, and two LIGO detectors \emph{and} Virgo 
active in O2. In each scenario, the blue and red histograms respectively refer to
the distributions $\mathcal{P}_{\rm BBH}(\log \mathcal{B}^{\rm ECO}_{\rm BBH})$ and
$\mathcal{P}_{\rm ECO}(\log \mathcal{B}^{\rm ECO}_{\rm BBH})$ defined in the previous section. 
The background distributions $\mathcal{P}_{\rm BBH}(\log \mathcal{B}^{\rm ECO}_{\rm BBH})$
are sufficiently well-behaved to allow for accurate Gaussian KDE approximations, with respect to which 
we calculate threshold values $\log \mathcal{B}_{\rm th}$ as in Eq.~(\ref{threshold}), for 
$p_{\rm th}$ corresponding to a significance of $5\sigma$. In each of the three cases,   
we then estimate the efficiency for $5\sigma$ detection of resonances by counting the fraction of 
foreground $\log \mathcal{B}^{\rm ECO}_{\rm BBH}$ samples that exceed $\log \mathcal{B}_{\rm th}$. 
This leads to efficiencies of, respectively, 17\%, 21\%, and 31\% for analyses in the
three data sets; as expected, the three-detector network with O2 sensitivity returns the highest 
efficiency. This indicates that O1 and O2 would have allowed for observations of
resonance-induced phase shifts with $\Delta\phi_{01}, \Delta\phi_{02} \lesssim 10$ rad.

To assess how low a phase shift can be detectable, we can look at subsets of 
foreground injections for which neither $\Delta\phi_{0i}$, $i = 1, 2$
exceeds some given value $\Delta\phi_{\rm max}$, and in each of the subsets determine 
efficiencies for $5\sigma$ detection of resonances. The results are shown in 
Fig.~\ref{fig:efficiency-trend}. 
We conclude that the chance of confidently detecting a resonance with 
\emph{e.g.}~$\Delta\phi_{0i} \leq 5$ is relatively low ($< 5\%$), and the signature of 
resonances mainly starts to be picked up from $\Delta\phi_{0i} \gtrsim 8$. 

\subsection{A note on parameter estimation}

Now let us turn to parameter estimation for resonance frequencies and phase shifts. 
Our Bayesian hypothesis $H_{\rm ECO}$ effectively assumes the presence of two resonances. 
Of course, in reality a binary coalescence involving ECOs may have zero instances of resonance in 
the detectors' sensitive frequency band, or only one, or more than two. 
Thus, though it is always possible to arrive at posterior 
density distributions for the two resonance frequencies 
$f_{01}$, $f_{02}$ and associated phase shifts 
$\Delta\phi_{01}$, $\Delta\phi_{02}$ through Eq.~(\ref{PE}), these should be taken with 
a grain of salt. Indeed, our real tool for assessing the presence of resonances is
$\log\mathcal{B}^{\rm ECO}_{\rm BBH}$ together with its background distribution 
$\mathcal{P}_{\rm BBH}(\log\mathcal{B}^{\rm ECO}_{\rm BBH})$. Nevertheless, for completeness
we show some representative example parameter estimation results for different cases.

First of all, Fig.~\ref{fig:PE_BBH} shows posterior densities for the case of a binary black hole injection.
No resonance frequencies are in band, and indeed the sampling puts most of the posterior
weight for $f_{01}$, $f_{02}$ at frequencies well above ISCO (where in this case 
$f_{\rm ISCO} = 97.4$ Hz). The distributions for $\Delta\phi_{01}$ and 
$\Delta\phi_{02}$ largely return the prior. 

In Fig.~\ref{fig:PE_2res}, we show posteriors for an ECO case with two resonance frequencies 
in band, for an event whose $\log\mathcal{B}^{\rm ECO}_{\rm BBH}$ is above
the $5\sigma$ threshold for detectability of resonances. 
The parameters related to resonances are estimated reasonably well. 

Finally, Fig.~\ref{fig:PE_1res} shows posteriors for an example ECO with only a \emph{single} resonance.
Though in this case parameter estimation cannot be fully reliable, the posteriors 
for $f_{01}$ and $\Delta\phi_{01}$ reasonably capture the true resonance frequency and 
phase shift. The posterior for $\Delta f_{02}$ only has support well above the ISCO frequency (here 
$f_{\rm ISCO} = 45.4$ Hz), 
reminiscent of the BBH case of Fig.~\ref{fig:PE_BBH}. However, we stress again that 
our ``detection statistic" $\log\mathcal{B}^{\rm ECO}_{\rm BBH}$ and its 
background distribution are what provide the means to establish the presence of resonances; 
and indeed, the log Bayes factor for this injection is comfortably above the $5\sigma$ threshold.

\section{Searching for resonances in GWTC-1 events}
\label{sec:gwtc-1}

Next we turn to analyzing the presumed binary black hole events of GWTC-1 \cite{GWTC1_open_data, Vallisneri_2015}. The main 
results are given by Fig.~\ref{fig:background-gwtc1} and Table \ref{tab:logB}. The 
Figure shows the values of $\log \mathcal{B}^{\rm ECO}_{\rm BBH}$ for the various events
in the three kinds of data sets. We also show again the background distributions 
$\mathcal{P}_{\rm BBH}(\log \mathcal{B}^{\rm ECO}_{\rm BBH})$, which the ``foreground"
log Bayes factors are clearly consistent with. In the Table, for each event we explicitly list 
log Bayes factors, as well as the false alarm probability with respect to the background
distribution. 

Two caveats are in order regarding the false alarm probabilities that we list. 
First, a larger number of BBH injections than the ones performed here will of course
result in a more accurate assessment of the background 
$\mathcal{P}_{\rm BBH}(\log \mathcal{B}^{\rm ECO}_{\rm BBH})$. Secondly, the
background will depend upon the distributions of masses and spins that were chosen for the 
injected BBH signals (specified in the previous section), but the astrophysical parameter distributions
for the population of heavy compact objects in the Universe are likely to differ from these.  
In the future one could use the \emph{measured} parameter distributions \cite{LIGOScientific:2018jsj}, 
whose accuracy will increase as more detections are made. That said, all of the
values of $\log \mathcal{B}^{\rm ECO}_{\rm BBH}$ that we obtain for individual events in 
GWTC-1 are anyway negative, thus favoring the BBH hypothesis. 

Though we find no evidence for the presence of resonances in any of the GWTC-1 events, for
completeness we show posterior density functions obtained for the resonance frequencies 
$f_{01}$, $f_{02}$ (Fig.~\ref{fig:freq}), and for the phase shifts $\Delta\phi_{01}$, 
$\Delta\phi_{02}$ (Fig.~\ref{fig:phase}). The posteriors for $f_{01}$, $f_{02}$ tend to be 
rather similar to the ones for the BBH injection of Fig.~\ref{fig:PE_BBH}, having most of their support 
at high frequencies, beyond ISCO. Also, the posteriors for $\Delta\phi_{01}$, $\Delta\phi_{02}$
are for the most part consistent with the priors.

\section{Summary and future directions}
\label{sec:conclusions}

Exotic compact objects 
may exhibit resonant excitations during inspiral, thereby 
taking away orbital energy from a binary system, leading to a speed-up of the orbital 
phase evolution relative to the binary black hole case. We have set up a Bayesian 
framework to look for such resonances under the assumption that they are of short 
duration, and that up to two resonant frequencies can be present in the part of 
the inspiral that is in the sensitive band of the Advanced LIGO and Advanced Virgo detectors. 
The associated model for the modification of the phase evolution allows one to compute 
a log Bayes factor $\log \mathcal{B}^{\rm ECO}_{\rm BBH}$ quantifying the ratio of evidences 
for the hypothesis that resonances occurred and the hypothesis that none were present. 

We calculated log Bayes factors for two sets of simulated signals embedded in data 
from the O1 and O2 observing runs, one in which the signals were from BBHs, and another
where resonances were present. Using the distribution of $\log \mathcal{B}^{\rm ECO}_{\rm BBH}$
from the former set as background and from the latter as foreground, we were able to conclude 
that the effect of resonance-induced phase shifts of $\Delta\phi_{01}, \Delta\phi_{02} \lesssim 10$ 
rad can be detectable at $5\sigma$ significance with an efficiency as large as $\sim 30$\%. 

We then turned to the presumed binary black hole events of GWTC-1. In all cases the  
$\log \mathcal{B}^{\rm ECO}_{\rm BBH}$ were found to be consistent with background, and moreover 
they were all negative, thus favoring the hypothesis that no resonances had occurred while the 
signals were in the detectors' sensitive frequency band. Posterior density functions
for resonance frequencies and induced phase shifts were consistent with these non-detections
of resonant excitations. 

Although so far we have found no evidence for resonances in binary black hole-like signals, it
is possible that this will happen in the future. In that case one will want to also characterize
the resonant excitations. The example framework presented here assumed two resonances in the 
ECO hypothesis $H_{\rm ECO}$. However, one could envisage Bayesian ranking within a list of 
ECO hypotheses $H^{(n)}_{\rm ECO}$ that assume there to be $n$ resonances in band, 
with $n = 1, 2, 3, \ldots$. Alternatively, one could have a single ECO hypothesis allowing for 
a variable number of resonances, with this number itself being sampled over. These further
improvements are left for future work.

Finally, with minor modifications our methodology could be used to search for resonant
r-modes in binary neutron star inspirals \cite{2007PhRvD..75d4001F,Balachandran:2007tu}. 
In that case the induced phase shifts are expected to be below detectable levels with 
existing instruments, but they may be in reach of more sensitive detectors in the
foreseeable future \cite{Poisson:2020eki}. Note that for r-modes the relevant parameters 
can be related to other properties of the neutron stars \cite{2007PhRvD..75d4001F}; for 
example the resonance frequencies are proportional to the spin frequencies, so that 
they need not be treated as completely free parameters. Setting up appropriate measurements
then deserves a separate, in-depth treatment; this is work in progress.

\section*{Acknowledgments}

We are grateful to Gideon Koekoek, Paolo Pani, Rob Tielemans, and Bert Vercnocke for very helpful 
discussions. 
Y.A.~acknowledges support from the International Research Experience for Undergraduates program of 
the University of Florida, which is funded by the U.S.~National Science Foundation. P.T.H.P., 
A.S., and C.V.D.B.~are supported by the research program of the Netherlands Organization 
for Scientific Research (NWO). 

This research has made use of data, software and/or web tools obtained from the Gravitational Wave Open Science Center (https://www.gw-openscience.org), a service of LIGO Laboratory, the LIGO Scientific Collaboration and the Virgo Collaboration. LIGO is funded by the U.S. National Science Foundation. Virgo is funded by the French Centre National de Recherche Scientifique (CNRS), the Italian Istituto Nazionale della Fisica Nucleare (INFN) and the Dutch Nikhef, with contributions by Polish and Hungarian institutes.

\bibliography{ref}  

\end{document}